# Susceptibility of Methicillin Resistant *Staphylococcus aureus* to Vancomycin using Liposomal Drug Delivery System.

Kiran Vishwasrao[1], Arjumanara Surti[2], Radha Srinivasan[1]

1: University of Mumbai, Mumbai 400 098, INDIA,

2: Sophia College, Bhulabhai Desai Road, Mumbai 400 026

Address for Correspondence:

Dr Radha Srinivasan, Associate Professor, Department of Physics, University of Mumbai, Mumbai, INDIA

*Abstract*

*Staphylococcus aureus* responsible for nosocomial infections is a significant threat to the public health. The increasing resistance of *S. aureus* to various antibiotics has drawn it to a prime focus for research on designing an appropriate drug delivery system. Emergence of Methicillin Resistant *Staphylococcus aureus* (MRSA) in 1961, necessitated the use of vancomycin "the drug of last resort" to treat these infections(6). Unfortunately, *S. aureus* has already started gaining resistances to vancomycin(5). Liposome encapsulation of drugs have been earlier shown to provide an efficient method of microbial inhibition in many cases(9). We have studied the effect of liposome encapsulated vancomycin on MRSA and evaluated the antibacterial activity of the liposome-entrapped drug in comparison to that of the free drug based on the minimum inhibitory concentration (MIC) of the drug. The MIC for liposomal vancomycin was found to be about half of that of free vancomycin. The growth response of MRSA showed that the liposomal vancomycin induced the culture to go into bacteriostatic state and phagocytic killing was enhanced. Administration of the antibiotic encapsulated in liposome thus was shown to greatly improve the drug delivery as well as the drug resistance caused by MRSA.

*Introduction*

Staphylococcal infections often occur exclusively in hospitals, playing on vulnerable patients, such as those with compromised immune systems or who had just undergone surgery. Earlier the introduction of penicillin in the 1940s stemmed staph infections in hospitals. However, staph strains quickly developed resistance to penicillin(23). Subsequently, methicillin, synthetic penicillin was developed but Staphylococcus quickly learned to evade it, as well. Methicillin resistant *Staphylococcus aureus* (MRSA) is now resistant to an entire class of penicillin-like antibiotics called beta-lactams. In 2002, physicians in the United States documented the first *S. aureus* strains resistant to the antibiotic, vancomycin, which had been one of the handful of antibiotics of last resort for use against *S. aureus*. Though it is feared that this could quickly become a major issue in antibiotic resistance, so far vancomycin- resistant strains are still rare(6, 23).

Vancomycin blocks the important substrates for cell wall synthesizing machinery, i.e. the D-alanyl-D-alanine residue (DDR) of the lipid II precursor in the cytoplasmic membrane of the bacterial cell(23). Thereby, it inhibits utilization of the substrates by glycosyltransferase (a cell wall synthesis enzyme) to produce the nascent peptidoglycan chain. However, besides the lipid II murein monomer precursors, which are the real targets of vancomycin, the cell wall peptidoglycan of a single *S. aureus* cell is known to posses about $6.0 \times 10^6$ DDRs to which vancomycin molecules could bind while penetrating the peptidoglycan layers and therefore is an inefficient drug in terms of maintaining an efficacious concentration around its real targets. Mechanistically the thickened cell wall is thought to limit the access of vancomycin to its target in MRSA strains(5).

Drug delivery is a major issue and delivery of free drug is often inefficient because the large size and low solubility of the drug in biofluid, lower penetration ability into the pathogens, development of drug resistance by the pathogen and other side effects(18). Encapsulation of the drug into liposomal vesicles can stabilize the drug and help in efficient delivery(14). The interaction of dioleoylphosphatidylcholine vesicles with murine tumor cells reported recently, which showed that small amounts of solute could be transferred from vesicles into cells(3). As shown by fluorescence microscopy and by fluorescence photobleaching recovery measurements, most of the vesicles were stably adsorbed on the cell surface with little incorporation of vesicle lipid into the cell membrane(3, 14). Liposomal encapsulation of tuberculostatic drugs such as isoniazid, pyrazinamide, rifampicin etc., in distearoyl phosphatidylcholine/cholesterol liposomes were shown to increase their therapeutic index(12). Soybean phospholipid (phosphatidylcholine) is a natural product, non toxic, biodegradable and non allergic(16) lipid. Liposome made up of Soybean phospholipid would be of advantage in drug delivery. The chances of drug resistance increases on the use of free drug since the pathogens are more likely to acquire resistance when exposed continuously to the drugs. Liposomes encapsulate the drug and thus prevent its exposure to the organism. They also reduce the time of exposure of the drug to the cell surface of the pathogen and hence do not allow the pathogens the time to develop resistance mechanisms. The liposome encapsulated drug thus may make the pathogens susceptible to low concentration of the drug(9, 11).

Liposomal drug delivery systems can contribute (i) to targeting of the drug to the infected tissues, (ii) in increasing intracellular antibiotic concentrations, and (iii) in reducing toxicity of potentially toxic drugs in certain cases(9, 17). As with most of the drug carriers, liposomes have been extensively used in an attempt to improve the selective delivery of antimicrobial agents and to reduce the side effects of the drug to some extent(6, 8, 11, 18). Encapsulation of ampicillin (Amp) in liposomes prepared with synthetic lecithins was shown to enhance its antibiotic activity against both Amp-sensitive and Amp-resistant *Escherichia coli*(22). Since the drug is encapsulated inside the vesicle, the recognition of the vesicle by the infected site is important for the targeting and the drug itself would have much less interaction with biofluid and associated components. This could reduce drug resistance caused due to delivery of free drug into the system to some extent(2).

To test this hypothesis in an *in vitro* set up, the culture of Methicillin Resistant *Staphylococcus aureus* (MRSA) was treated with the liposomal vancomycin preparation. This liposomal vancomycin preparation was then tested on the MRSA culture in comparison to free vancomycin to determine the minimum inhibitory concentrations(11, 18) of both the forms of the drug. The growth response of MRSA on exposure to the free drug and the liposomal drug delivery systems were monitored using growth curves over a period of six hours. The effect of liposomal vancomycin on phagocytosis was also studied. Lastly, the efficacy of the liposomal vancomycin on storage of the liposomal preparation was checked for a period of three months.

## *Materials and Methods*

**Characterization of Methicillin resistant *Staphylococcus aureus*:** For all inoculation purposes the culture density was adjusted to 0.1 OD at 530 nm in Muller-Hinton Broth (MHB) containing 2% NaCl(9, 20). Antibiotic susceptibility test – Kirby Bauer method to reconfirm that the procured *Staphylococcus aureus* culture was Vancomycin intermediate was carried out using Muller-Hinton Agar containing 2% NaCl. The antibiotic disc used was vancomycin (10 μg) disc. Well isolated colony of MRSA was added to 3mL of MH- broth incubated at $37^0C$ for 3 hours density adjusted to 0.5 OD at 530 nm. The plates were incubated at $37^0C$ for 24 hours. Bioassay using vancomycin drug VANSAFE- CP by VBH Medisciences Ltd. used in the experiment was carried out to indicate that the test organism is likely to respond to therapy.

**Preparation of liposome:** Soybean phospholipid was prepared by using reported method(9, 19). The purified Soybean phospholipid was dissolved in 1ml of chloroform and then subjected to nitrogen gas to evaporate the chloroform from the vial. A thin layer of phospholipid is formed on the inside wall of the vial. This was then vacuum dried at $37^0C$ / 3 hrs. Rehydration was carried out by adding 1ml of distilled water for plain liposome. In case of drug loaded liposome: 50mg/ml of vancomycin dissolved in distilled water was added to the vacuum dried vial. This was then subjected to 10 cycles of freeze and thaw (liquid nitrogen – $40^0$ C water bath). To obtain small unilamellar vesicles, the solution was passed through a stainless steel extrusion device (Unimetrics, Illinois, USA), first with a 0.2 µm membrane (Costar Scientific, MA) and then with a 0.05 µm membrane. A PerkinElmer UV-Vis-NIR spectrometer was used for absorption spectra of vancomycin and vancomycin entrapped liposomes to determine the concentration of vancomycin in the samples. The hydrodynamic radius was determined using Dynamic light scattering (DynaPro 99).

**Susceptibility test procedure:** The antimicrobial activity of vancomycin –loaded liposomes was determined in comparison with that of the free drug, with Minimum inhibitory concentration using the standard broth microdilution assay(7, 9, 21). Minimum inhibitory concentration (MIC), is the lowest concentration of an antimicrobial agent that will inhibit the visible growth of a microorganism after overnight incubation. Stock solutions of liposomal vancomycin (117µg/ml) and free vancomycin (50 µg/ml) were used. Appropriate controls were maintained. Plates were incubated at 37°C for 24 hours. Each assay was repeated three times.

**Growth response:** growth curve was carried out to check the effect of subleathal- MIC concentrations of liposomal vancomycin and free vancomycin in comparison to normal curve (in absence of any form of drug). 100µl of culture (0.1 OD) was added to 4.9ml of Muller-Hinton Broth (MHB). A control i.e. without the drug was used. Total volume was 10ml. Readings were taken at intervals of 30 minutes for 6 hours.

**Phagocytosis:** Phagocytosis in presences of liposomal vancomycin was studied using phagocytosis experiment(15). To separate leukocyte from the whole blood, 1.5 mL of 3% dextran in saline is added to 5mL of heparinised blood. Then leukocyte rich plasma is separated from red blood cells by centrifuging it at 100 rpm/ 10min. For phagocytosis assay three tubes are used labeled as Control, Test 1and Test2. In each tube 0.25 ml AB serum, 0.25 mL Hanks balanced salt solution, 0.25 mL *S. aureus and* 0.25 mL Leukocyte (except in control) is added and incubated at $37^0C$/1hr. In Test2 after a 30 min/$37^0C$ 0.25mL liposomal vancomycin is added. After incubation period 1mL cold Trypan blue dye is added to each tube. Centrifuged at 1500 rpm/ 10min. supernatant is discarded and the cells suspended in the remaining volume, mixed thoroughly. Examined using a hemocytometer, dead cells blue and live cells yellow using light microscope.

**Efficiency of liposomal drug on storage:** The effect of material of storage namely plastic and glass and effect of temperature ($4^0C$)(2) on the antibacterial efficiency of the vancomycin loaded liposome (liposomal drug) was carried out by storing the colloidal suspension in plastic and glass vials under Refrigeration ($4^0C$) for a period of 3 months. At interval of one month the efficiency testing was carried out by determining the MIC- using micro-broth dilution method(7, 9) (method mentioned above).

## *Results & Discussion*

**Characterization of Methicillin Resistant *Staphylococcus aureus*:** The antibiotic susceptibility assay was carried out using Kirby Bauer method(21) to reconfirm its sensitivity to vancomycin. These zone size of 20mm for vancomycin disc agreed with that expected for *S. aureus* ATCC 25923 (17-22mm) as reported earlier. The bioassay results indicated that the MRSA culture showed 'intermediate susceptibility' to the vancomycin drug VANSAFE-CP which was planed to be used as the drug for the experiment, indicating that the test organism is likely to respond to therapy with this drug.

**Determination of concentration of vancomycin entrapped liposome:** The UV-visible absorption spectra of vancomycin shows characteristic peak at 280nm (Figure 1). The extinction coefficient of the drug at 280nm was determined to be 8743/mol/cm. The plain liposome prepared from soybean phospholipid did not show any appreciable absorbance at 280nm but the absorption spectra of the drug encapsulated liposome showed the absorption band at 280 nm which can only be due to the presence of vancomycin. The concentration of vancomycin entrapped in the liposome was found to be 11730μg/mL as determined from the UV-visible absorption spectra.

**Characterization of liposome:** Hydrodynamic radius of plain liposomes and liposomal vancomycin was determined using dynamic light scattering (DLS) technique. Since the liposomes are present in a colloidal suspension they are in continuous Brownian motion. To determine the hydrodynamic radius of the liposomes, a monomodal distribution of size was considered in the analyses of the DLS data. The hydrodynamic radius for plain liposome and vancomycin loaded liposome were found to be 123 nm and 157 nm respectively (figure 2). The size distribution of the liposomes observed in the present case agree well with earlier reports(13).

**Susceptibility test procedure:** Minimum inhibitory concentration (MIC) of liposomal vancomycin as compared to free vancomycin was determined using micro broth dilution method(7, 9, 21). Standard vancomycin powder is expected to give MIC values in the range of 0.5μg/mL to 2.0μg/mL (NCCLS)(1, 7). The assay of the inhibitory effect based on the determination of turbidity as a signature of presences of growth showed that the MIC of free vancomycin for MRSA in our preparation was 1.5 -1.6μg/mL, which is in the range given in the NCCLS chart.

Controls with plain liposomes were maintained and it was observed that the plain liposomes did not show any inhibitory effect on MRSA growth. Thus ensuring that the inhibitory effect if any observed was solely due to the drug and the delivery system. The assay showed for vancomycin entrapped liposomes was 0.8μg/mL (figure 3). Thus the MIC for vancomycin decreased to half when the drug was delivered using liposomal encapsulation. In other words, using liposomal drug delivery system the concentration of vancomycin required to inhibit the same volume of culture was half the concentration required when the same vancomycin is delivered in the free form.

Thus strongly indicating that liposomal drug delivery system is a very effective system for drug delivery(14, 21). Since low concentration of drug is required to inhibit the culture the chances of resistance induced due to exposure to high concentration of drug can be combated.

**Growth curve analysis:** To check the effect of liposomal drug in comparison with the free drug on the growth cycle of MRSA growth curve studies were carried out. Half the MIC i.e. sub-lethal concentration of the free vancomycin and liposomal vancomycin was used to study growth curve response. The growth curve of MRSA showed exponential growth with time under normal conditions of growth i.e. in the absence of any form of drug (neither free vancomycin nor liposomal vancomycin) the culture showed a short lag phase and soon entered the log phase of growth within one hour (figure 4). With free vancomycin the culture showed a longer lag phase and entered the log phase only after two hours and with time the curve showed a plateau indicating arrest of cell multiplication. This indicated that the free vancomycin has considerable inhibitory effect on MRSA. When the MRSA was exposed to liposomal vancomycin the culture showed no growth for three hours and negligible growth after six hours. This indicated that the cell growth was arrested as soon as the cells were exposed to liposomal vancomycin.

**Phagocytosis:** Phagocytosis in presences of liposomal vancomycin was studied using phagocytosis assay. The phagocytosis assay showed that in the presence of the drug-loaded-liposome only dead (blue) cells were present in the phagocyte whereas in the control both the live (yellow) and the dead (blue) cells were present. This indicated that the drug was delivered efficiently into the phagocyte killing all the *S. aureus* cells enhancing the bactericidal activity of the phagosome. This confirms that the liposomes increase the efficiency of drug delivery.

**Effect of storage on liposome:** The effect of storage on liposomal vancomycin was studied with respect to time, temperature and material of storage. The MIC value remained consistent for a period of 3 months at refrigeration temperature ($4^0$C)(2). When efficacy was tested on storage in plastic and glass vials no change in MIC values was observed for vancomycin entrapped liposomes i.e. the MIC value remained consistent for a period of 3 months. Thus indicating that the liposomal suspension can be stored efficiently at $4^0$C in a refrigerator, in either plastic or glass vials without the fear of losing its antibacterial activity at least for a period of 3 months.

## *Conclusion*

The liposomal vancomycin was extremely effective in inhibiting the MRSA culture at lower concentration of vancomycin. The MIC values decreased from 1.6µg/mL to 0.8µg/mL when the drug in entrapped into soybean phospholipid. Thus indicating that the drug concentration required to inhibit the same volume culture reduced to half its concentration when delivered using liposomes. The growth curve studies indicated that when MRSA is exposed to liposomal vancomycin it immediately enters into a static state with negligible growth over a period of six hours. Whereas, when exposed to free drug the culture enters into the log phase within two hours and multiplies steadily for the next three hours after which the curve starts to plateau indicating that the culture has stoped multiplying. Thus the culture is inhibited at a faster rate when the the drug is entrapped in liposomes as compared to free drug. Phagocytosed cells can be killed more efficiently and faster using liposomal vancomycin.

Liposomal drug delivery system is more effective in inhibiting and eventually killing bacteria than the traditional free drug delivery system. The drug is delivered more efficiently using liposome at lower concentrations and at a faster rate i.e. short time interval to reach its target site in a concentrated form. Whereas when free drug is delivered the drug gets diluted by the time it reaches its real target i.e. longer time period are required with higher concentration of drug to inhibit the same volume of culture. By the time (longer time) the free drug reaches its real targets the drug is lost at various stages – cell wall, the drug is affected by enzymes in the cytoplasm (enzymatic action) hence (diluting) reducing the final effective concentration of the drug delivered. Liposomal drug delivery system seems to be the best option to cure infections and overcome drug resistant.

## *Acknowledgement:*

This work was partly carried out by Ms Priya Putta and was supported by Prof S. Majumdar, Department of Chemical Physics, Tata Institute of Fundamental Research, Mumbai.

## Declaration of Interest:

The authors report no declarations of interest.

## References

1. **Andrews, J. M.** 2001. Determination of minimum inhibitory concentrations J Antimicrob Chemother **48:**5-16.
2. **Beaulac, C., S. Sachetelli, and J. Lagace.** 1998. In-vitro bactericidal efficacy of sub-MIC concentrations of liposome-encapsulated antibiotic against gram-negative and gram-positive bacteria. J Antimicrob Chemother **41:**35-41.
3. **Blumenthal, R., E. Ralston, P. Dragsten, L. D. Leserman, and J. N. Weinstein.** 1982. Lipid vesicle-cell interactions: analysis of a model for transfer of contents from adsorbed vesicles to cells. Membr Biochem **4:**283-303.
4. **Brown, D. F., D. I. Edwards, P. M. Hawkey, D. Morrison, G. L. Ridgway, K. J. Towner, and M. W. Wren.** 2005. Guidelines for the laboratory diagnosis and susceptibility testing of methicillin-resistant Staphylococcus aureus (MRSA). J Antimicrob Chemother **56:**1000-18.
5. **Cui, L., A. Iwamoto, J. Q. Lian, H. M. Neoh, T. Maruyama, Y. Horikawa, and K. Hiramatsu.** 2006. Novel mechanism of antibiotic resistance originating in vancomycin-intermediate Staphylococcus aureus. Antimicrob Agents Chemother **50:**428-38.
6. **Davenport, R. J.** 2007. A shifting threat Development of antibiotic resistance and spread outside of hospitals weighs on researchers efforts to tame Staphylococcus, Infection Research: Perspectives.
7. **Espinel-Ingroff, A., J. L. Rodriguez-Tudela, and J. V. Martinez-Suarez.** 1995. Comparison of two alternative microdilution procedures with the National Committee for Clinical Laboratory Standards reference macrodilution method M27-P for in vitro testing of fluconazole-resistant and -susceptible isolates of Candida albicans. J Clin Microbiol **33:**3154-8.
8. **Fresta, M., E. Wehrli, and G. Puglisi.** 1995. Enhanced therapeutic effect of cytidine-5'-diphosphate choline when associated with GM1 containing small liposomes as demonstrated in a rat ischemia model. Pharm Res **12:**1769-74.
9. **Furneri, P. M., M. Fresta, G. Puglisi, and G. Tempera.** 2000. Ofloxacin-loaded liposomes: in vitro activity and drug accumulation in bacteria. Antimicrob Agents Chemother **44:**2458-64.
10. **Geha, D. J., J. R. Uhl, C. A. Gustaferro, and D. H. Persingl.** 1994. Multiplex PCR for identification of Methicillin-Resistant Staphylococci in the clinical laboratory. Journal Of Clinical Microbiology **32:**1768-1772.
11. **Hospenthal, D. R., A. L. Rogers, and E. S. Beneke.** 1989. Effect of attachment of anticandidal antibody to the surfaces of liposomes encapsulating amphotericin B in the treatment of murine candidiasis. Antimicrob Agents Chemother **33:**16-8.
12. **Justo, O. R., and A. M. Moraes.** 2003. Incorporation of antibiotics in liposomes designed for tuberculosis therapy by inhalation. Drug Deliv **10:**201-7.

13. **Justo, O. R., and A. M. Moraes.** 2005. Kanamycin incorporation in lipid vesicles prepared by ethanol injection designed for tuberculosis treatment. J Pharm Pharmacol **57:**23-30.
14. **Kozubek, A., J. Gubernator, E. Przeworska, and M. Stasiuk.** 2000. Liposomal drug delivery, a novel approach: PLARosomes. Acta Biochim Pol **47:**639-49.
15. **Kubica, M., K. Guzik, J. Koziel, M. Zarebski, W. Richter, B. Gajkowska, A. Golda, A. Maciag-Gudowska, K. Brix, L. Shaw, T. Foster, and J. Potempa.** 2008. A potential new pathway for Staphylococcus aureus dissemination: the silent survival of S. aureus phagocytosed by human monocyte-derived macrophages. PLoS ONE **3:**e1409.
16. **Lautenschlager, H.** 2006. Liposomes, p. 155-163. *In* A. O. Barel, M. Paye, and H. I. Maibach (ed.), Handbook of Cosmetic Science and Technology. CRC Press Taylor & Francis Group, Boca Raton.
17. **Lee, R. J.** 2006. Liposomal delivery as a mechanism to enhance synergism between anticancer drugs. Mol Cancer Ther **5:**1639-40.
18. **Levine, D. P.** 2006. Vancomycin: a history. Clin Infect Dis **42 Suppl 1:**S5-12.
19. **McAllister, S. M., H. O. Alpar, and M. R. Brown.** 1999. Antimicrobial properties of liposomal polymyxin B. J Antimicrob Chemother **43:**203-10.
20. **Mugabe, C., M. Halwani, A. O. Azghani, R. M. Lafrenie, and A. Omri.** 2006. Mechanism of enhanced activity of liposome-entrapped aminoglycosides against resistant strains of Pseudomonas aeruginosa. Antimicrob Agents Chemother **50:**2016-22.
21. **Pulimood, T. B., M. K. Lalitha, M. V. Jesudason, R. Pandian, J. Selwyn, and T. J. John.** 1996. The spectrum of antimicrobial resistance among methicillin resistant Staphylococcus aureus (MRSA) in a tertiary care centre in India. Indian J Med Res **103:**212-5.
22. **Sammour, O. A., and H. M. Hassan.** 1996. Enhancement of the antibacterial activity of ampicillin by liposome encapsulation. Drug Deliv **3:**273-278.
23. **Scherl, A., P. Francois, Y. Charbonnier, J. M. Deshusses, T. Koessler, A. Huyghe, M. Bento, J. Stahl-Zeng, A. Fischer, A. Masselot, A. Vaezzadeh, F. Galle, A. Renzoni, P. Vaudaux, D. Lew, C. G. Zimmermann-Ivol, P. A. Binz, J. C. Sanchez, D. F. Hochstrasser, and J. Schrenzel.** 2006. Exploring glycopeptide-resistance in Staphylococcus aureus: a combined proteomics and transcriptomics approach for the identification of resistance-related markers. BMC Genomics **7:**296.

## *Figure Captions:*

[1] UV-visible absorption spectra for vancomycin entrapped in liposomes (Lipo+Vanco: solid lines), free vancomycin (Free Vanc: dashed lines) and plain liposomes (Only Lipo: dotted lines).

[2] Dynamic light scattering of (A) Soybean liposome and (B) Liposome entrapped vancomycin. The results were analysed by mono-modal distribution to obtain the mean radius of the vescicles.

[3] Plot of minimum inhibitory concentration (MIC) of vancomycin for *S. aureus* with free vancomycin (empty squares), liposome entrapped vancomycin (solid triangles). Solid line represent positive control of the culture and dotted line is negative control (media without *S. aureus*).

[4] Growth curves of *S. aureus* in presence of free vancomycin (solid circles), liposome entrapped vancomycin (solid triangles) and control i.e., without vancomycin/liposome (solid squares)

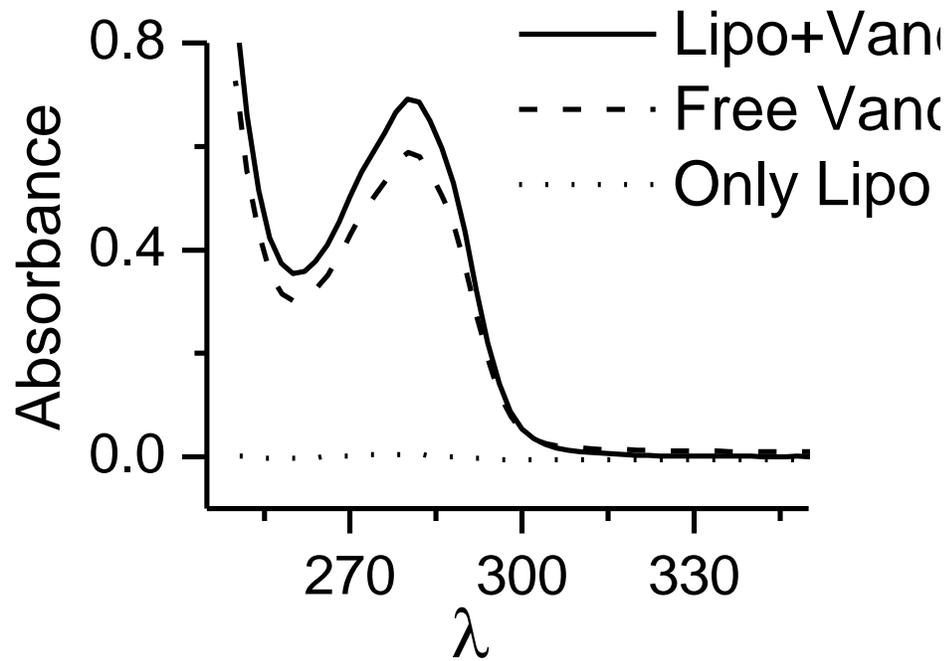

Figure 1: UV-visible absorption spectra for vancomycin entrapped in liposomes (Lipo+Vanco: solid lines), free vancomycin (Free Vanc: dashed lines) and plain liposomes (Only Lipo: dotted lines).

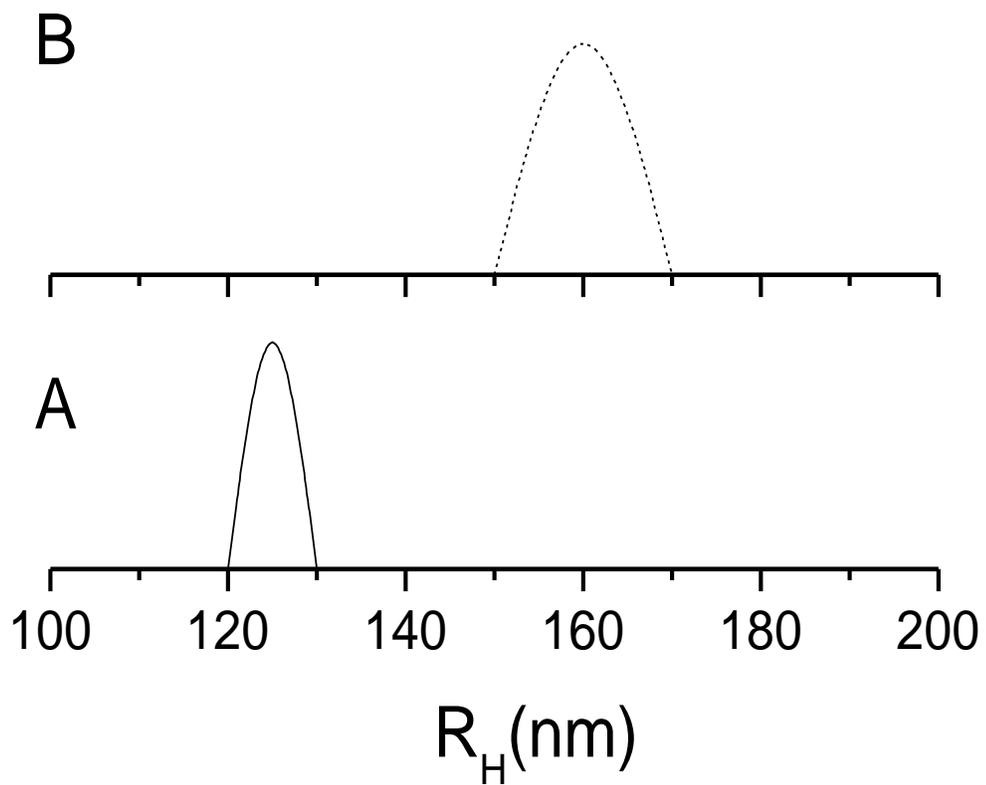

Figure 2: Dynamic light scattering of (A) Soybean liposome and (B) Liposome entrapped vancomycin. The results were analysed by mono-modal distribution to obtain the mean radius of the vescicles.

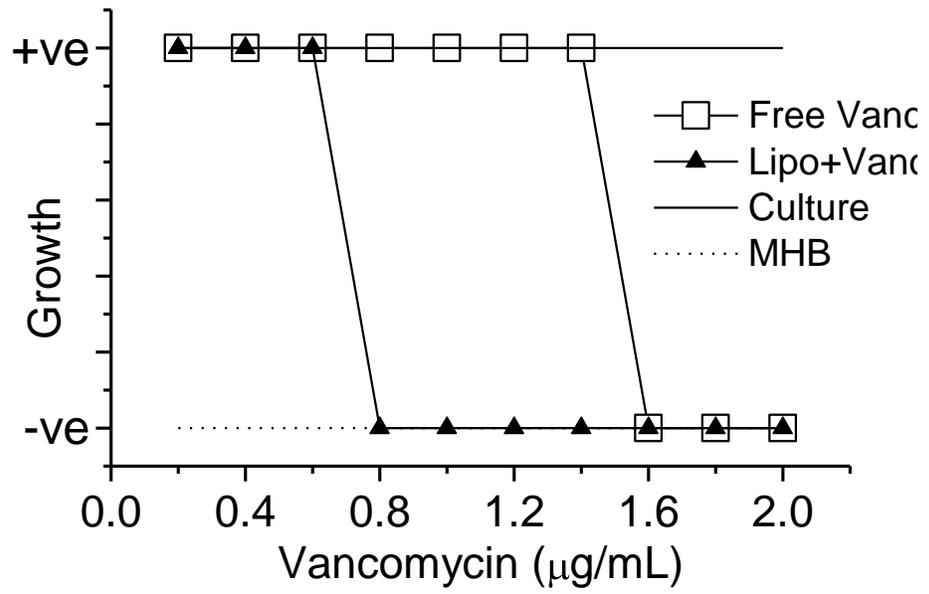

Figure 3: Plot of minimum inhibitory concentration (MIC) of vancomycin for *S. aureus* with free vancomycin (empty squares), liposome entrapped vancomycin (solid triangles). Solid line represent positive control of the culture and dotted line is negative control (media without *S. aureus*).

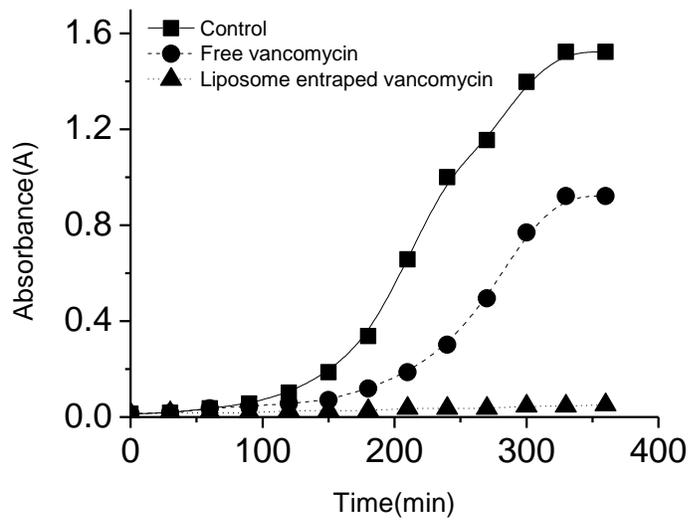

Figure 4: Growth curves of *S. aureus* in presence of free vancomycin (solid circles), liposome entrapped vancomycin (solid triangles) and control i.e., without vancomycin/liposome (solid squares)